\documentclass{article}

\usepackage{multirow}
\usepackage[table]{xcolor}
\usepackage{pifont}
\usepackage[switch]{lineno}  
\usepackage{etoolbox}
\usepackage[table]{xcolor}

\usepackage[normalem]{ulem}
\usepackage{booktabs}
\usepackage{multirow}
\usepackage{array}
\usepackage{makecell}
\usepackage{graphicx}
\usepackage{threeparttable}
\usepackage{bm}
\definecolor{cmp}{HTML}{E8F5E9}
\definecolor{base}{HTML}{E3F2FD}
\definecolor{abl}{HTML}{E8EAF6}
\definecolor{enc}{HTML}{EDE7F6}
\definecolor{rob}{HTML}{FFF3E0}
\definecolor{cmp}{HTML}{E8F5E9}
\definecolor{base}{HTML}{E3F2FD}
\definecolor{abl}{HTML}{E8EAF6}
\definecolor{enc}{HTML}{EDE7F6}
\definecolor{rob}{HTML}{FFF3E0}
\definecolor{best}{gray}{0.85}
\definecolor{second}{gray}{0.93}
\definecolor{RColor}{HTML}{0805CD}

 \usepackage[preprint]{icml2026}

\usepackage{amsmath}
\usepackage{amssymb}
\usepackage{mathtools}
\usepackage{amsthm}

\usepackage[capitalize,noabbrev]{cleveref}

\theoremstyle{plain}

\theoremstyle{definition}

\theoremstyle{remark}

\usepackage[textsize=tiny]{todonotes}

\icmltitlerunning{Unifying Speech Editing Detection and Content Localization via Prior-Enhanced Audio LLMs}

\begin{document}
	
	\twocolumn[
	\icmltitle{Unifying Speech Editing Detection and Content Localization via Prior-Enhanced Audio LLMs}
	
	
	
	\icmlsetsymbol{equal}{*}
	
%
%

	\begin{icmlauthorlist}
		\icmlauthor{Jun Xue}{equal,whu}
		\icmlauthor{Yi Chai}{equal,whu}
		\icmlauthor{Yanzhen Ren}{whu}
		\icmlauthor{Jinshen He}{ind}
		\icmlauthor{Zhiqiang Tang}{ahu}
		\icmlauthor{Zhuolin Yi}{whu}
		\icmlauthor{Yihuan Huang}{whu}
		\icmlauthor{Yuankun Xie}{cuc}
		\icmlauthor{Yujie Chen}{buaa}
	\end{icmlauthorlist}
	
	\icmlaffiliation{whu}{Key Laboratory of Aerospace Information Security and Trusted Computing, Ministry of Education; School of Cyber Science and Engineering, Wuhan University, Wuhan, China}
	\icmlaffiliation{ind}{Independent Researcher}
	\icmlaffiliation{ahu}{School of Computer Science and Technology, Anhui University, Hefei, China}
	\icmlaffiliation{cuc}{Communication University of China, Beijing, China}
	\icmlaffiliation{buaa}{Beihang University, Beijing, China}
	
	\icmlcorrespondingauthor{Yanzhen Ren}{renyz@whu.edu.cn}

\icmlkeywords{Speech Editing Detection, Content Localization, Audio LLMs, Deepfake Detection}
	
	\vskip 0.3in
	]

	
	
	\printAffiliationsAndNotice{}  

\begin{abstract}
Existing speech editing detection (SED) datasets are predominantly constructed using manual splicing or limited editing operations, resulting in restricted diversity and poor coverage of realistic editing scenarios. Meanwhile, current SED methods rely heavily on frame-level supervision to detect observable acoustic anomalies, which fundamentally limits their ability to handle deletion-type edits, where the manipulated content is entirely absent from the signal.
To address these challenges, we present a unified framework that bridges speech editing detection and content localization through a generative formulation based on Audio Large Language Models (Audio LLMs). We first introduce AiEdit\footnote{\url{https://huggingface.co/datasets/JunXueTech/AiEdit}}, a large-scale bilingual dataset (approximately 140 hours) that covers addition, deletion, and modification operations using state-of-the-art end-to-end speech editing systems, providing a more realistic benchmark for modern threats.
Building upon this, we reformulate SED as a structured text generation task, enabling joint reasoning over edit type identification, and content localization. To enhance the grounding of generative models in acoustic evidence, we propose a prior-enhanced prompting strategy that injects word-level probabilistic cues derived from a frame-level detector. Furthermore, we introduce an acoustic consistency-aware loss that explicitly enforces the separation between normal and anomalous acoustic representations in the latent space.
Experimental results demonstrate that the proposed approach consistently outperforms existing methods across both detection and localization tasks.

\end{abstract}

\section{Introduction}
\label{sec:intro}

Recent advances in large-scale data and deep learning have significantly improved the quality and naturalness of speech generation systems. While these technologies greatly enhance content creation efficiency, they also introduce serious risks to personal privacy and societal security. Audio deepfakes can be broadly categorized into two types. The first is fully synthetic speech, which generates complete utterances using Text-to-Speech (TTS)~\cite{chen2025f5} or Voice Conversion (VC)~\cite{yao2025stablevc} techniques, thereby altering the global distribution of speech signals. The second is speech editing~\cite{fluentspeech}, which manipulates local segments of an existing utterance through addition, deletion, or modification, enabling semantic changes while largely preserving speaker identity and acoustic context. Compared to fully synthetic speech, speech editing is more covert, as it only modifies partial regions, making it harder to detect and potentially more harmful in real-world scenarios such as misinformation and impersonation.

Existing speech editing techniques can be roughly divided into two categories. Early approaches rely on rule-based manual editing~\cite{partialspoof}, which typically involves VAD-based segmentation, segment generation, and subsequent splicing. These methods often introduce perceptible discontinuities at editing boundaries, allowing detection models to rely on low-level acoustic artifacts. In contrast, recent end-to-end neural speech editing methods~\cite{partialedit} generate edited segments in a semantically conditioned manner, achieving high consistency with surrounding prosody and context. As a result, editing traces become significantly less perceptible, leading to more realistic and challenging detection scenarios. However, existing speech editing detection (SED) datasets are still predominantly constructed based on manual splicing or limited editing types (e.g., modification only), which restricts their ability to reflect the diversity and realism of modern editing techniques.

From a modeling perspective, current SED methods~\cite{martíndoñas2022vicomtechaudiodeepfakedetection,tdl,cfprf,bam} primarily adopt frame-level binary supervision to detect local acoustic anomalies. While effective for addition and modification, such approaches fundamentally rely on the presence of observable artifacts in the signal. This assumption becomes problematic for deletion-type edits, where the manipulated content is entirely absent from the observed waveform. As a result, deletion is inherently difficult to localize based on direct acoustic evidence, which limits the effectiveness of conventional frame-level detection methods in fine-grained speech editing analysis.

To address these challenges, we propose a unified framework that jointly tackles speech editing detection and content localization through a generative formulation. First, we construct \textbf{AiEdit}, a bilingual speech editing dataset (approximately 140 hours), covering addition, deletion, and modification operations generated by multiple state-of-the-art end-to-end editing systems. This dataset provides a more realistic and comprehensive benchmark for evaluating modern editing threats.
Building upon this, we introduce a prior-enhanced Audio Large Language Model (\textbf{PELM}) that reformulates SED as a structured text generation task, enabling unified reasoning over detection, edit type identification, and content localization. To improve the grounding of generative reasoning in acoustic evidence, we design a prior-enhanced prompting mechanism that incorporates word-level probabilistic cues derived from a frame-level detector. Furthermore, to mitigate the tendency of Audio LLMs to over-rely on semantic information~\cite{röttger2024xstesttestsuiteidentifying}, we introduce an acoustic consistency-aware loss that explicitly enforces the separation between normal and anomalous acoustic representations in the latent space.

The main contributions of this work are summarized as follows:
\begin{itemize}
	\item To our best knowledge, AiEdit is the first bilingual speech editing dataset that covers addition, deletion, and modification operations, providing a more comprehensive benchmark for modern speech editing threats.
	\item We propose a unified generative framework based on Audio LLMs for joint speech editing detection and content localization, and introduce a prior-enhanced prompting strategy and an acoustic consistency-aware loss to better align acoustic evidence with generative reasoning.
	\item Extensive experiments demonstrate that the proposed method consistently outperforms existing approaches on both detection and localization tasks.
\end{itemize}

\section{Motivation}
\subsection{Threat methods}
\label{sec:TM}

\begin{figure}[t]
	\centering
	\includegraphics[width=0.9\linewidth]{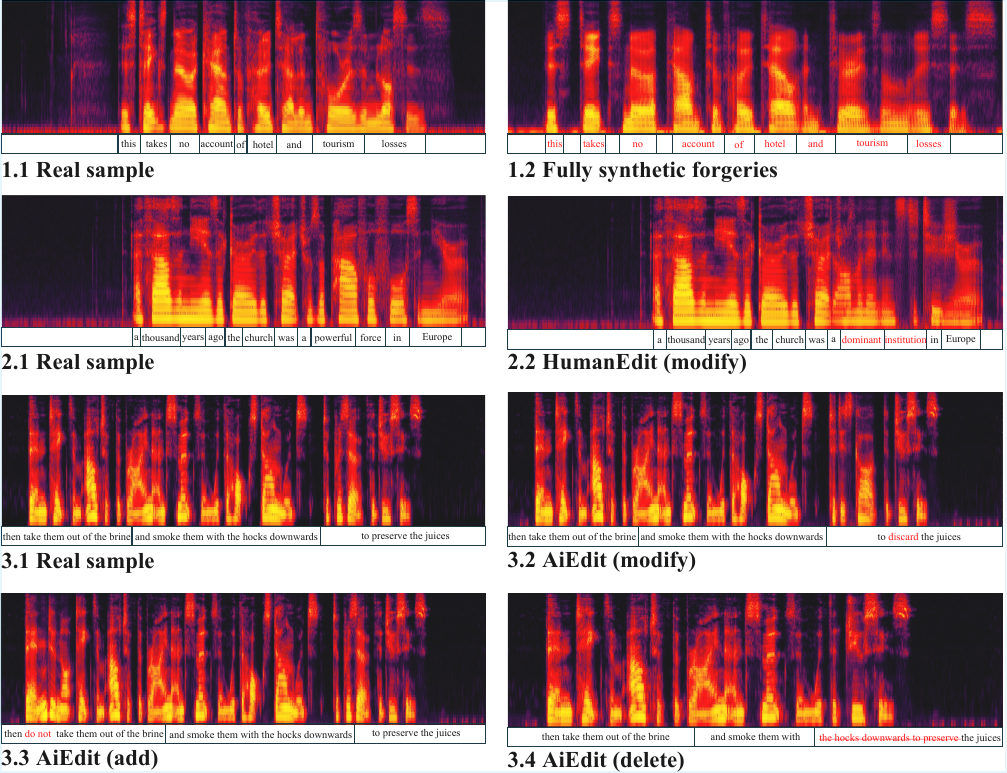}
	\caption{Spectrogram comparison across different editing samples.}
	\label{fig:dataset_cmp}
\end{figure}

\begin{table*}[t]
	\centering
	\caption{Statistical comparison and attribute overview of partial spoofing datasets. Note that ``-'' indicates the information is not publicly available or applicable.}
	\label{tab:dataset_comparision}
	\resizebox{\textwidth}{!}{%
		\begin{tabular}{lcccccccc}
			\toprule
			\textbf{Dataset} & \textbf{Year} & \textbf{Datasource} & \textbf{Language} & \textbf{Operations} & \textbf{Access} & \multicolumn{2}{c}{\textbf{\#Utterances}} & \textbf{MOS} \\
			\cmidrule(lr){7-8}
			& & & & & & \textbf{Real} & \textbf{Fake} & \\
			\midrule
			PartialSpoof \cite{partialspoof}   & 2021 & ASVspoof 2019 LA & English & modify & Public & 12,483 & 108,978 & $3.41 \pm 0.21$ \\
			HAD \cite{had}                    & 2021 & AISHELL-3 & Chinese & modify & Public & 53,612 & 53,612 & $3.43 \pm 0.17$ \\
			ADD2022Track2 \cite{add2022}      & 2022 & AISHELL-3 & Chinese & modify & Public & 5,319 & 1,052 & -- \\
			Psynd \cite{psynd}                & 2022 & LibriTTS & English & modify & Restrict & -- & -- & -- \\
			LAV-DF \cite{lavdf}               & 2022 & Voxceleb2 & English & modify & Public & 36,431 & 99,873 & -- \\
			ADD2023Track2 \cite{add2023}      & 2023 & AISHELL-3 & Chinese & modify & Public & 55,467 & 63,831 & -- \\
			PartialEdit \cite{partialedit}    & 2025 & Voxceleb2 & English & modify & Partially Public & -- & 43,358 & $3.41 \pm 0.21$ \\
			LlamaPartialSpoof \cite{llamapartialspoof} & 2025 & LibriTTS & English & modify & Public & 10,573 & 32,194 & -- \\
			\midrule
			\textbf{AiEdit (Ours) }         & 2026 & Libriheavy/Chineselip & Chinese/English & add/delete/modify & Public & 7,760 & 51,794 & $3.77 \pm 0.36$ \\
			\bottomrule
		\end{tabular}%
	}
\end{table*}

\noindent\textbf{Fully synthetic forgery:}
Early research on fake speech detection primarily focused on fully synthetic attacks, which aim to spoof entire utterances to compromise automatic speaker verification (ASV) systems, as exemplified by the ASVspoof and ADD challenge series. As illustrated in Figs.~\ref{fig:dataset_cmp} (1.1 and 1.2), although the generated speech exhibits high perceptual quality, its global speaking characteristics are altered, enabling reliable discrimination from bona fide speech based on distributional differences.

\noindent\textbf{Human-assisted editing:}
To preserve most characteristics of the original speaker while manipulating local semantics, human-assisted editing (HumanEdit) based on the `cut-and-paste'' paradigm has been proposed, with PartialSpoof~\footnote{In this paper, the term `HumanEdit'' refers to the PartialSpoof dataset.}\cite{partialspoof} as a representative work. This approach synthesizes local segments via TTS or VC and then splices them into bona fide speech. As shown in Figs.~\ref{fig:dataset_cmp} (2.1 and 2.2), such methods introduce noticeable acoustic artifacts at editing boundaries and often result in unnatural transitions or semantic inconsistencies. Consequently, detection models can still rely on low-level acoustic artifacts for discrimination.

\noindent\textbf{End-to-end neural speech editing:}
Recent end-to-end neural speech editing (AiEdit) methods enable more flexible and realistic manipulations, including addition, deletion, and modification. Under semantic guidance, these models generate edited segments that are highly consistent with the surrounding context in terms of prosody and speaking style. As illustrated in Figs.~\ref{fig:dataset_cmp} (3.1--3.4), such edits exhibit smooth transitions and minimal artifacts, making them significantly harder to detect using conventional artifact-based methods.

Compared with fully synthetic and manually edited speech, modern neural speech editing introduces more subtle and diverse manipulation patterns, especially with the inclusion of deletion operations. This evolution fundamentally shifts the detection paradigm from artifact-based discrimination to more semantic-aware analysis, highlighting the need for models that can handle diverse editing types and subtle inconsistencies.

\subsection{Audio LLMs}
\label{sec:ALLM}

\begin{figure}[tbp]
	\centering
	\includegraphics[width=\linewidth]{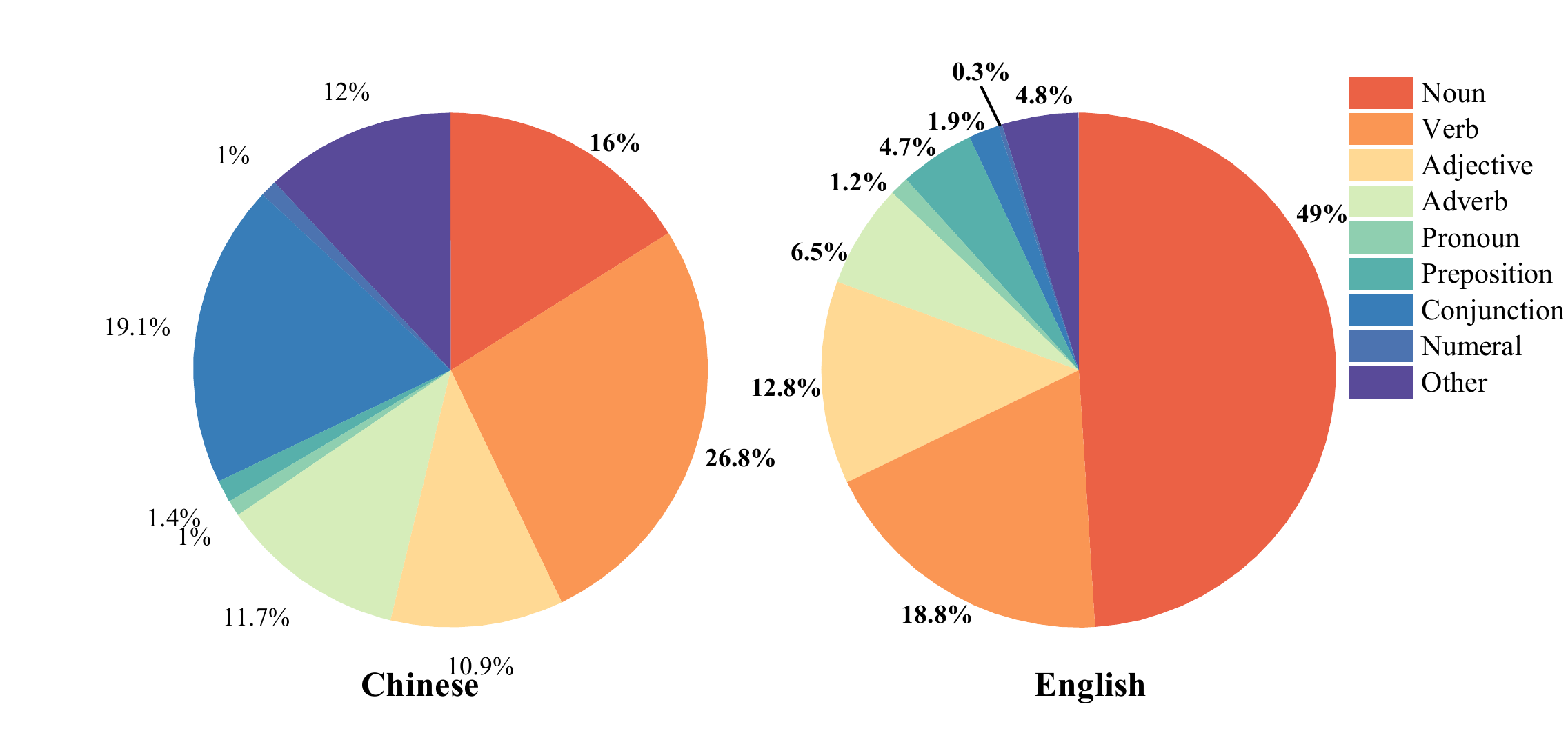}
	\caption{Distribution of Part-of-Speech (POS) tags for edited words. The pie charts illustrate the proportion of different syntactic categories targeted for editing in the Chinese (left) and English (right) subsets of our dataset.}
	\label{fig:pos_distribution}
\end{figure}

In recent years, Audio LLMs have shown strong capabilities in audio understanding and generation tasks, and have demonstrated promising performance in fully synthetic speech detection~\cite{allm4add}. Unlike conventional frame-level models, Audio LLMs can naturally model speech in a generative manner, making them suitable for tasks that require semantic-level reasoning and flexible output representations. This makes them particularly suitable for scenarios where acoustic evidence alone is insufficient.

Such properties are beneficial for SED, where the goal is not only to determine whether an utterance is edited, but also to identify the edit type and localize the edited content. In particular, for deletion-type edits, the manipulated content is absent from the observed signal, making it difficult to model using frame-level supervision. In contrast, Audio LLMs provide a more flexible way to reason about such cases.

However, directly applying Audio LLMs to this task remains challenging. During fine-tuning, these models tend to rely heavily on semantic information~\cite{wang2025pay} and may overlook subtle acoustic inconsistencies introduced by speech editing. As a result, the model may produce semantically plausible outputs without sufficiently grounding its predictions in acoustic evidence.

To address these limitations, we introduce an acoustic consistency-aware loss that explicitly enforces the modeling of local acoustic anomalies in the feature space. In addition, we incorporate word-level prior information derived from a frame-level detector to provide auxiliary acoustic cues, improving the alignment between semantic reasoning and acoustic evidence.

\section{The AiEdit dataset}

With the advancement of speech editing techniques, existing speech editing detection datasets still exhibit limitations in editing types and construction methods. As shown in Table~\ref{tab:dataset_comparision}, most existing datasets focus on modification-only operations and are typically built using manual splicing or limited generation approaches, which fail to reflect the more natural and diverse editing scenarios introduced by modern end-to-end speech editing methods.
To address these limitations, we construct a bilingual speech editing dataset, AiEdit, which covers three types of operations: addition (Add), deletion (Delete), and modification (Modify), and leverages multiple end-to-end speech editing models to generate edited speech. In terms of data split, the test set further includes editing methods that are unseen during training to evaluate model generalization.
In the following, we describe the text construction and speech generation processes, as well as the overall data distribution and split settings.

\subsection{Dataset construction}	

\begin{figure}[tbp]
	\centering
	\includegraphics[width=\linewidth]{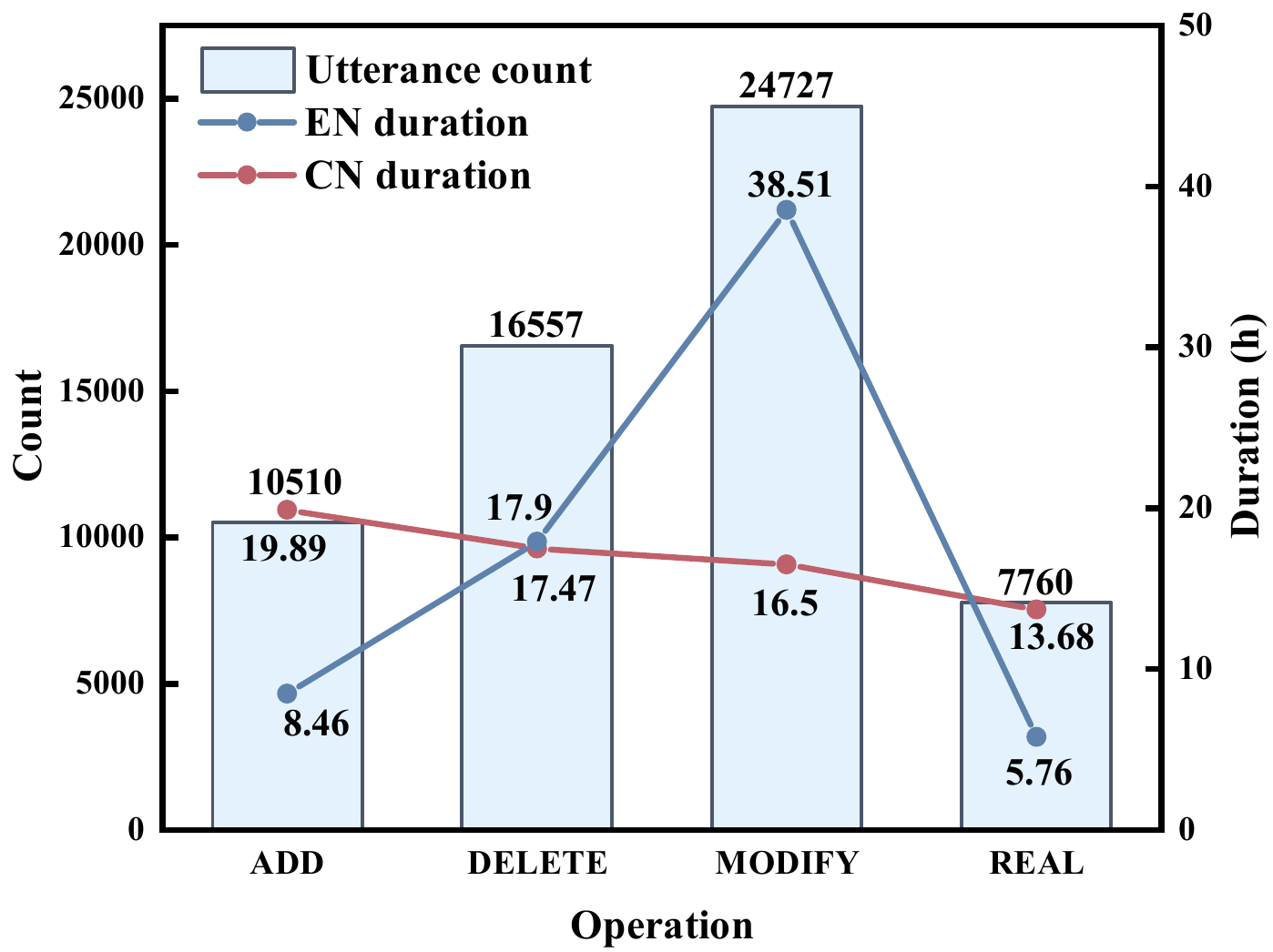}
	\caption{Statistical overview of the dataset composition. The bar chart (left axis) displays the total sample count for each operation type (Add, Delete, Modify) and real speech. The line plot (right axis) illustrates the total audio duration in hours for the English (blue) and Chinese (red) subsets.}
	\label{fig:duration}
\end{figure}

\textbf{Text generation:} during the construction of the AiEdit dataset, we first model speech editing at the textual level to ensure controllability and consistency in semantic manipulation. To this end, we strictly define three basic types of editing operations: addition (Add), deletion (Delete), and modification (Modify), and impose constraints on each operation to guarantee atomicity and semantic validity.

Specifically, the Add operation inserts semantically relevant content into the original text and is restricted to a single continuous insertion point. The Delete operation removes a segment from the original sentence; to preserve semantic integrity, deletions at sentence boundaries are strictly prohibited, and only a single continuous deletion region is allowed. The Modify operation replaces a segment of the original text with new content, requiring the replacement to be similar in length and semantically consistent.

For text generation, we employ the large language model Qwen3-max\footnote{\url{https://qwen.ai/blog?id=qwen3-max}} to perform semantically guided editing on the original text. To ensure data quality, we adopt an effective quality control pipeline. First, generated samples are validated through text differencing by performing word-level alignment between the source and edited texts, ensuring compliance with predefined editing constraints (e.g., single-region edits and non-boundary deletions). In addition, beyond automated validation, we further incorporate manual inspection to provide complementary quality assurance and ensure the overall reliability of the dataset.

To further assess the semantic diversity of the edited content, we analyze the part-of-speech (POS) distribution of the tampered regions. As illustrated in Fig.~\ref{fig:pos_distribution}, by leveraging a large language model for semantically guided editing, AiEdit covers a wide range of POS categories, including nouns, verbs, and adjectives, resulting in diverse editing patterns. Specifically, different languages exhibit distinct distributions. In the English subset, edits are dominated by nouns (49\%), followed by verbs (18.8\%) and adjectives (12.8\%), which is consistent with the prominence of noun phrases in English. In contrast, the Chinese subset shows a more balanced distribution, where verbs account for the largest proportion (26.8\%), followed by nouns (16\%) and adverbs (11.7\%). These observations indicate that the constructed dataset captures diverse semantic editing patterns while maintaining reasonable linguistic structures.

\begin{table}[t]
	\centering
	\begin{threeparttable}
		\caption{Statistics of sample quantities across different dataset splits and operation types}
		\label{tab:sample_distribution}
		\setlength{\tabcolsep}{2pt}
		\begin{tabular}{lccccc}
			\toprule
			& \textbf{Real} & \textbf{ADD} & \textbf{DELETE} & \textbf{MODIFY} & \textbf{Edit Methods} \\
			\midrule
			\textbf{Train} & 1993 & 1753  & 2790   & 4792   & E1-E4 \\
			\textbf{Val}   & 397  & 357   & 561    & 941    & E1-E4 \\
			\textbf{Test}  & 5370 & 8411  & 13206  & 18983  & E1-E6 \\
			\bottomrule
		\end{tabular}
		
		\begin{tablenotes}
			\small
			\item \textbf{Methods:} $E_1$: FluentSpeech, $E_2$: Ming, $E_3$: PlayDiffusion, $E_4$: VoiceCraft, $E_5$: A3T, $E_6$: SSR.
		\end{tablenotes}
	\end{threeparttable}
\end{table}

\textbf{Speech editing:}
In the speech generation stage, we employ multiple state-of-the-art end-to-end speech editing models to synthesize edited audio, to improve the diversity and realism of the dataset. Specifically, we adopt several representative approaches, including A3T\footnote{\url{https://github.com/richardbaihe/a3t}}~\cite{a3t}, FluentSpeech\footnote{\url{https://github.com/Zain-Jiang/Speech-Editing-Toolkit}}~\cite{fluentspeech}, Ming\footnote{\url{https://github.com/inclusionAI/Ming-UniAudio}}~\cite{ming}, PlayDiffusion\footnote{\url{https://github.com/playht/PlayDiffusion}}~\cite{playdiff}, SSR\footnote{\url{https://github.com/WangHelin1997/SSR-Speech}}~\cite{ssr}, and VoiceCraft\footnote{\url{https://github.com/jasonppy/VoiceCraft}}~\cite{voicecraft}.
These models cover diverse modeling paradigms, including autoregressive generation, diffusion-based methods, and speech reconstruction frameworks, enabling support for multiple editing operations such as addition, deletion, and modification.

\begin{figure*}[t]
	\centering
	\includegraphics[width=0.9\linewidth]{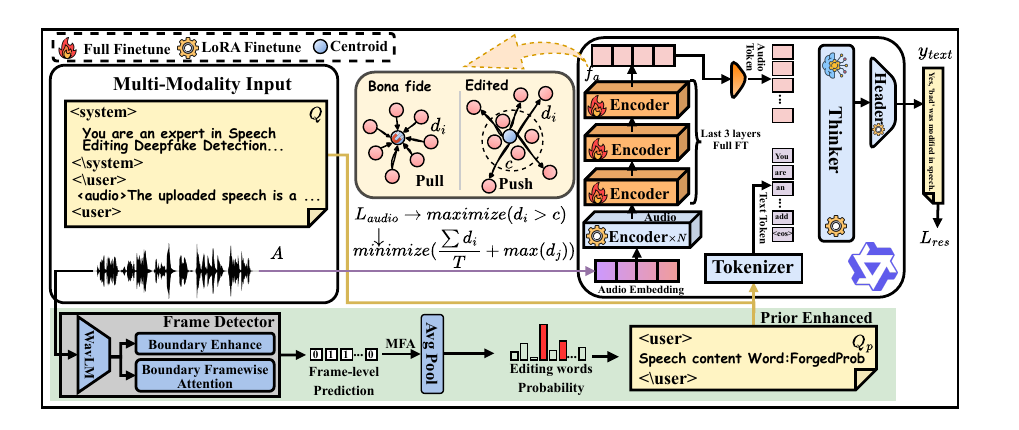}
	\caption{Overview of the PELM architecture, including prior-enhanced multi-modality input construction, Audio LLMs-based reasoning, and centroid clustering-based training objective.}
	\label{fig:method}
\end{figure*}

\subsection{Dataset statistics}

This section presents a statistical overview of the AiEdit dataset in terms of sample distribution, audio duration, and data splits.

\textbf{Overall distribution.}
As shown in Fig.~\ref{fig:duration}, the dataset consists of three editing operations (Add, Delete, Modify) and real speech (Real). The Modify category contains 24,727 samples, the Delete category contains 16,557 samples, the Add category contains 10,510 samples, and the Real category contains 7,760 samples. 

In terms of audio duration, for the English subset, the total durations of Add, Delete, Modify, and Real are 8.46, 17.9, 38.51, and 5.76 hours, respectively. For the Chinese subset, the corresponding durations are 19.89, 17.47, 16.5, and 13.68 hours.

\textbf{Data split.}
As reported in Table~\ref{tab:sample_distribution}, the dataset is divided into training, validation, and test sets. The training set contains 1,993 real samples and 9,335 edited samples, while the validation set includes 397 real samples and 1,859 edited samples. The test set consists of 5,370 real samples and 40,600 edited samples.
The training and validation sets comprise samples generated by editing methods E1--E4, while the test set includes not only E1--E4 but also unseen editing attacks E5 and E6.

\section{Method}

\subsection{Task definition}

We formulate speech editing analysis as a structured text generation task. Given an input speech utterance $A$ and a natural language instruction prompt $Q$, the model generates a textual response $y$ that jointly indicates whether the utterance is edited, the edit type, and the edited content.

To enable standardized evaluation, we impose strict output format constraints. For bona fide inputs, the model outputs a fixed statement: 
``No evidence of speech editing was detected.'' 
Otherwise, the output follows a structured template:
``Yes, \textless exact words\textgreater\ were \textless Type\textgreater\ in the speech.'',
where $\textless$Type$\textgreater \in \{\textit{added, deleted, modified}\}$ specifies the editing operation. For the deletion type, <exact words> refers to the missing semantic units inferred by the model's generative reasoning based on the surrounding context.

\subsection{Unified Audio LLMs framework}

The overall architecture of PELM is illustrated in Fig.~\ref{fig:method}. We adopt an Audio LLM as the backbone to jointly model acoustic input and textual instructions. Given the input audio $A$ and instruction prompt $Q$, the model generates a structured response:
\begin{equation}
	y = M(A, Q),
\end{equation}
where $M$ denotes the Audio LLM.

This formulation enables unified modeling of detection and localization, allowing the model to reason over both acoustic signals and semantic context.

\subsection{Prior-enhanced prompting}

To provide auxiliary acoustic evidence and improve grounding, we introduce a word-level probabilistic prior derived from a frame-level detector.

Specifically, given an audio input $A$, a pre-trained frame-level detector $D$ produces frame-wise editing probabilities. Using forced alignment, the frame-level probabilities are aggregated within each word boundary to obtain a word-level probability sequence $p$. The aggregation is implemented via average pooling over aligned frames.

The resulting sequence $p$ is then encoded into a structured prior prompt $Q_p$, which is concatenated with the original instruction prompt and fed into the Audio LLM. The final generation process is defined as:
\begin{equation}
	y = M(A, Q, Q_p).
\end{equation}

This design injects low-level acoustic cues into the generative reasoning process, improving alignment between semantic prediction and acoustic evidence.

\subsection{Acoustic consistency-aware loss}

To mitigate the model's over-reliance on semantic information, we introduce an acoustic consistency-aware loss based on centroid clustering to explicitly enforce the modeling of local acoustic inconsistencies.

Let $f_a \in \mathbb{R}^{L \times d}$ denote the hidden representation sequence corresponding to audio tokens, where $L$ is the number of audio tokens and $d$ is the feature dimension. The centroid is computed as:
\begin{equation}
	c = \frac{1}{L} \sum_{i=1}^{L} f_a^i,
\end{equation}
and the cosine distance between each token representation and the centroid is defined as:
\begin{equation}
	d_i = 1 - \cos(f_a^i, c).
\end{equation}

We design an asymmetric objective for bona fide and edited speech:
\begin{equation}
	\small
	\mathcal{L}_{audio} =
	\begin{cases}
		\frac{1}{L} \sum_{i=1}^{L} d_i + \max_i d_i, & A \in \mathcal{B}, \\[6pt]
		\frac{1}{|S_{topk}|} \sum\limits_{i \in S_{topk}} \mathrm{ReLU}(m - d_i), & A \in \mathcal{E},
	\end{cases}
\end{equation}
where $\mathcal{B}$ and $\mathcal{E}$ denote bona fide and edited speech, respectively, $S_{topk}$ is the set of top-$K\%$ tokens with the largest distances, which focuses on the most salient anomalous regions, and $m$ is a margin.

For bona fide speech, we enforce a \textit{cohesion constraint} that encourages all audio-token representations to remain compact around the centroid. For edited speech, we focus on tokens with large deviations and apply a margin-based constraint to encourage the presence of anomalous patterns.

\subsection{Training objective}

The overall training objective combines the standard cross-entropy loss for text generation with the proposed acoustic consistency-aware loss:
\begin{equation}
	\mathcal{L}_{total} = \mathcal{L}_{ce} + \lambda \cdot \mathcal{L}_{audio},
\end{equation}
where
\begin{equation}
	\mathcal{L}_{ce} = -\frac{1}{T} \sum_{t=1}^{T} \log P(y_t \mid y_{<t}, A, Q, Q_p),
\end{equation}
and $\lambda$ is a balancing coefficient.

The cross-entropy loss supervises structured text generation, while the acoustic consistency-aware loss enhances the model's ability to capture fine-grained acoustic inconsistencies.

\section{Experiments}
\label{sec:exp}

\subsection{Experimental setup}

\begin{figure}[t]
	\centering
	\includegraphics[width=\linewidth]{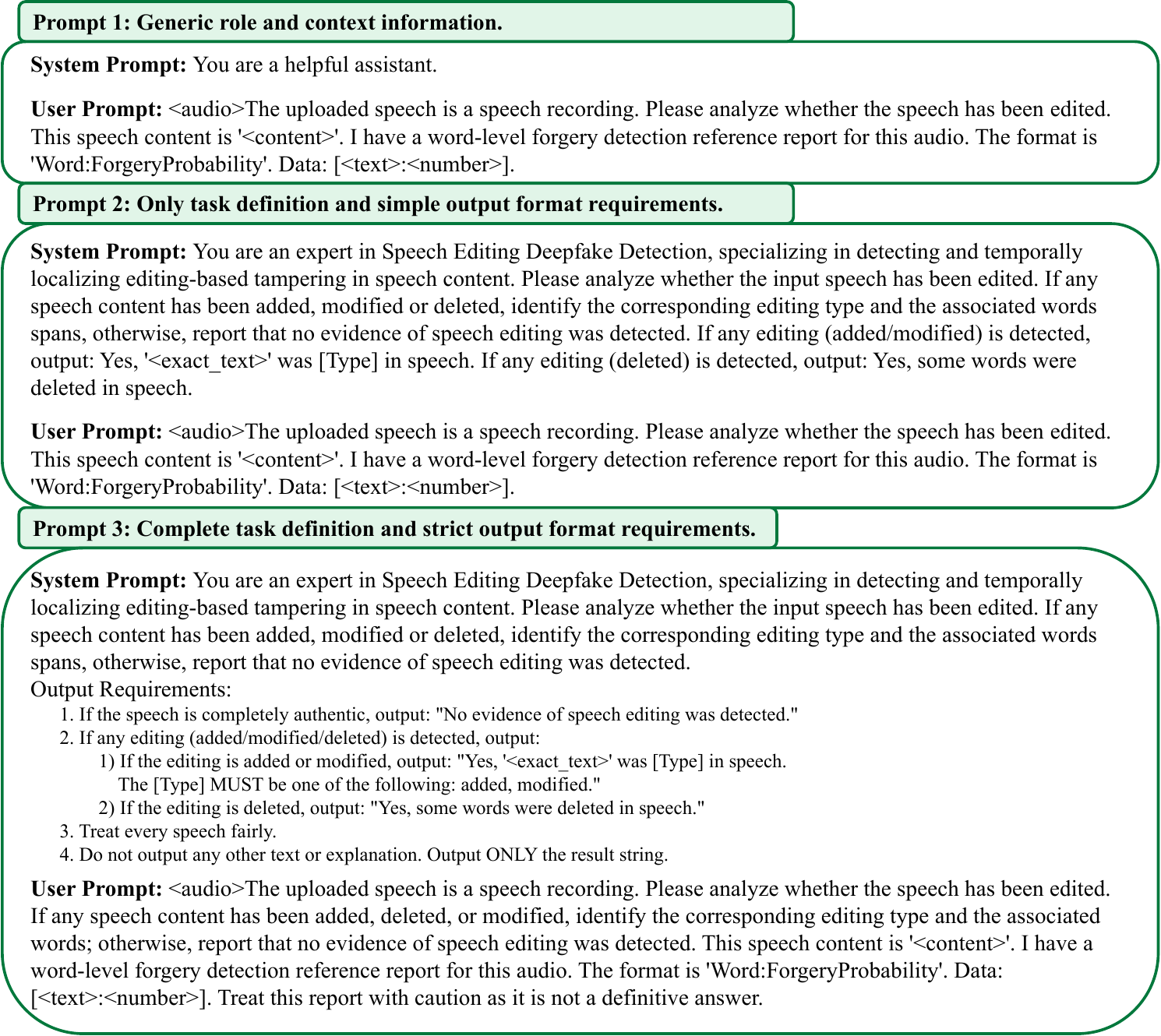}
	\caption{The details of different prompt.}	
	\label{fig:prompt1}
\end{figure}

In this work, we conduct experiments based on several Audio LLMs, including Qwen2.5-Omni-3B~\cite{Qwen2.5-Omni}, Qwen2.5-Omni-7B~\cite{Qwen2.5-Omni}, and Qwen2-Audio-7B~\cite{chu2024qwen2}. We design a series of prompts~\ref{fig:prompt1} with increasing levels of specificity to guide the model, ranging from basic contextual instructions (Prompt 1) to fully constrained task formulations with strict output formats (Prompt 3). Unless otherwise specified, Prompt 3 is used by default. A frame-level detector is first employed to estimate the probability of tampering for each frame. The resulting frame-level probabilities are then aggregated into word-level probabilities using forced alignment (MFA~\cite{mcauliffe2017montreal}) based on word boundaries, which are further constructed as prior prompts and injected into the Audio LLMs.  For model training, we adopt the ms-swift framework and employ a LoRA-based fine-tuning strategy for efficiency, while fully fine-tuning the last three layers of the audio encoder and the projection layer. Detailed hyperparameter settings are provided in Table~\ref{tab:hyperparameters}. All experiments are conducted on a single NVIDIA RTX 5880 GPU.

For comparison, we reproduce several representative conventional speech editing detection methods, including TDL\footnote{\url{https://github.com/xieyuankun/TDL-ADD}}, CFPRF\footnote{\url{https://github.com/ItzJuny/CFPRF}}, AGO\footnote{\url{https://github.com/Little-dingding/ATGO}}, and BAM\footnote{\url{https://github.com/media-sec-lab/BAM}}. These methods are primarily based on frame-level or boundary-aware modeling, aiming to detect editing traces through local acoustic anomalies or discontinuities. For fair comparison, all baselines are implemented based on their official repositories under the same data splits and evaluation settings.

In terms of datasets, we adopt PartialSpoof~\cite{partialspoof} as a representative dataset for manually constructed editing scenarios, referred to as \textbf{HumanEdit}, while the proposed \textbf{AiEdit} dataset is used to characterize end-to-end neural speech editing. AiEdit covers three types of editing operations: addition, deletion, and modification, where deletion operations do not have explicit corresponding segments in the observed speech signal. It is worth noting that conventional methods are typically trained with frame-level binary supervision and rely on detecting anomalous segments explicitly present in the audio, making them unsuitable for handling deletion-type edits. Based on this limitation, deletion samples are excluded from the AiEdit test set when evaluating conventional methods.

\subsection{Evaluation metrics}

\textbf{Detection.}
We first evaluate model performance at the utterance level, where the task is to determine whether a given speech sample has been edited. This is formulated as a binary classification problem. We report several standard metrics, including Equal Error Rate (EER), F1 score, and Area Under the ROC Curve (AUC). EER reflects the trade-off between false acceptance and false rejection rates, while F1 and AUC provide complementary perspectives on classification performance under different decision thresholds.

\begin{table}[t]
	\centering
	\caption{Hyperparameter settings for fine-tuning the Default model.}
	\label{tab:hyperparameters}
	
	\begin{tabular}{lc}
		\toprule
		\textbf{Hyperparameters} & \textbf{Value} \\
		\midrule
		Base Model          & Qwen-2.5-Omni-3B \\
		Training Epochs     & 5 \\
		Global Batch Size   & 8 \\
		Max Sequence Length & 2048 \\
		
		Optimizer           & AdamW \\
		Learning Rate       & $1 \times 10^{-5}$ \\
		LR Scheduler        & Cosine \\
		Warmup Ratio        & 0.2 \\
		Weight Decay        & 0.1 \\
		
		LoRA Rank ($r$)     & 32 \\
		LoRA Alpha ($\alpha$) & 64 \\
		LoRA Dropout        & 0.05 \\
		
		$ \lambda$   & 0.5 \\
		margin & 0.9 \\
		topk & 0.1 \\
		\bottomrule
	\end{tabular}
\end{table}

\noindent\textbf{Content localization.}
We further evaluate the ability of the model to identify edited content at the semantic level. Specifically, the model is required to localize edited units, defined as words for English and characters for Chinese. For unified evaluation, we adopt Word Error Rate (WER) as the metric for both languages.

The WER is computed following the standard formulation used in automatic speech recognition:
\begin{equation}
	\text{WER} = \frac{S + D + I}{N},
\end{equation}
where $S$, $D$, and $I$ denote the number of substitution, deletion, and insertion errors, respectively, and $N$ is the total number of words (or characters) in the reference sequence.

In this work, the reference sequence corresponds to the ground-truth edited content, while the hypothesis sequence is obtained from the model-generated output. A lower WER indicates more accurate localization of edited content.

\begin{table*}[t!]
	\centering
	\begin{threeparttable}
		\caption{Model Performance Comparison on HumanEdit and AiEdit, and their average (Pool). Best results are \textbf{bolded} and second-best results are \underline{underlined}.}
		\label{tab:main_1}
		\scriptsize
		\setlength{\tabcolsep}{3pt}
		\renewcommand{\arraystretch}{1.02}
		
		\begin{tabular}{l ccc c ccc ccc c ccc c}
			\toprule
			\multirow{3}{*}{\textbf{Model}} 
			& \multicolumn{4}{c}{\textbf{HumanEdit}} 
			& \multicolumn{7}{c}{\textbf{AiEdit}} 
			& \multicolumn{4}{c}{\textbf{Pool}} \\
			\cmidrule(lr){2-5} \cmidrule(lr){6-12} \cmidrule(lr){13-16}
			
			& \multicolumn{3}{c}{Det.} & \multicolumn{1}{c}{Loc.} 
			& \multicolumn{3}{c}{Det.\tnote{*}} & \multicolumn{3}{c}{Det.\tnote{**}} & \multicolumn{1}{c}{Loc.}
			& \multicolumn{3}{c}{Det.\tnote{***}} & \multicolumn{1}{c}{Loc.} \\
			\cmidrule(lr){2-4} \cmidrule(lr){5-5} 
			\cmidrule(lr){6-8} \cmidrule(lr){9-11} \cmidrule(lr){12-12}
			\cmidrule(lr){13-15} \cmidrule(lr){16-16}
			
			& ACC$\uparrow$ &  F1$\uparrow$ & EER$\downarrow$ 
			& WER$\downarrow$ 
			& ACC$\uparrow$ &  F1$\uparrow$ & EER$\downarrow$ 
			& ACC$\uparrow$ &  F1$\uparrow$ & EER$\downarrow$ 
			& WER$\downarrow$ 
			& ACC$\uparrow$ &  F1$\uparrow$ & EER$\downarrow$ 
			& WER$\downarrow$ \\
			\midrule
			
			TDL~\cite{tdl}
			& 96.25 & 98.06 & 14.69 
			& 15.02
			
			& 34.95 & 43.86 & 49.95  						& -- & -- & -- 
			
			& 9.81
			& 65.60 & 70.96 & 32.32 
			& 12.42\\ 
			
			CFPRF~\cite{cfprf}
			& 97.69 & 98.71 & 5.81 
			& 27.27
			
			& 65.54 & 74.89 & 31.44  						& -- & -- & --
			
			& 10.63
			& 81.62 & 86.80 & 18.63 
			&18.95\\
			
			AGO~\cite{ago} 
			& 97.76 & 98.75 & 4.17 
			& \cellcolor{second}\underline{13.42} 		
			& 44.93 & 35.48 & 28.03  	& -- & -- & --
			& 8.72
			& 71.35 & 67.12 & 16.10 
			& 14.96\\
			
			BAM~\cite{bam}
			& \cellcolor{second}\underline{99.56} & \cellcolor{second}\underline{99.76} & \cellcolor{second}\underline{0.63} 
			& 14.10 		
			& \cellcolor{second}\underline{78.48} & \cellcolor{second}\underline{85.87} & \cellcolor{second}\underline{11.66} 
			& -- & -- & --
			& \cellcolor{second}\underline{6.53}
			& \cellcolor{second}\underline{89.02} & \cellcolor{second}\underline{92.82} & \cellcolor{second}\underline{6.15} 
			&\cellcolor{second}\underline{10.13} \\
			
			\textbf{PELM (Ours)} 
			& \cellcolor{best}\textbf{99.62} & \cellcolor{best}\textbf{99.78} & \cellcolor{best}\textbf{0.55} 
			& \cellcolor{best}\textbf{9.17} 			& \cellcolor{best}\textbf{96.01} & \cellcolor{best}\textbf{95.98}& \cellcolor{best}\textbf{10.14}
			& \cellcolor{best}\textbf{95.2} & \cellcolor{best}\textbf{97.19} & \cellcolor{best}\textbf{8.37} 
			
			& \cellcolor{best}\textbf{2.72}
			& \cellcolor{best}\textbf{97.82} & \cellcolor{best}\textbf{97.88} & \cellcolor{best}\textbf{5.35} 
			& \cellcolor{best}\textbf{6.82}\\
			\bottomrule
		\end{tabular}
		\begin{tablenotes}
			\item*denotes detection excluding the deletion type. **denotes detection including the deletion type. *** denotes the average results on HumanEdit and *.
		\end{tablenotes}
	\end{threeparttable}
\end{table*}

\subsection{Overall performace}
Table~\ref{tab:main_1} summarizes the detection and content localization performance across different methods on both HumanEdit and AiEdit datasets.

For the detection task, all methods achieve strong performance on the HumanEdit dataset, which can be attributed to the presence of relatively explicit acoustic or boundary cues that are easier to capture. However, on the AiEdit dataset, the performance of conventional methods degrades noticeably. Under the setting excluding deletion operations (Det*), TDL, CFPRF, and AGO exhibit significantly lower performance compared to HumanEdit, while BAM maintains relatively better detection performance but still shows overall degradation. It is important to note that conventional methods are based on frame-level binary supervision and rely on anomalous segments explicitly present in the audio, making them unsuitable for detecting deletion-type edits. As a result, they cannot be evaluated under the full setting including deletion operations (Det**). In contrast, the proposed PELM maintains stable performance under both settings and achieves strong detection results even when deletion is included.

For the content localization task, conventional methods generally perform worse, especially on the AiEdit dataset where WER values remain relatively high. In comparison, PELM achieves the lowest WER on both datasets, further reducing it to 2.72\% on AiEdit, demonstrating its advantage in semantic-level localization. Finally, in terms of the overall average results (Pool), PELM achieves the best performance across both detection and localization metrics, further validating its robustness and generalization across different data distributions.

\begin{figure}[t]
	\centering
	\includegraphics[width=0.9\linewidth]{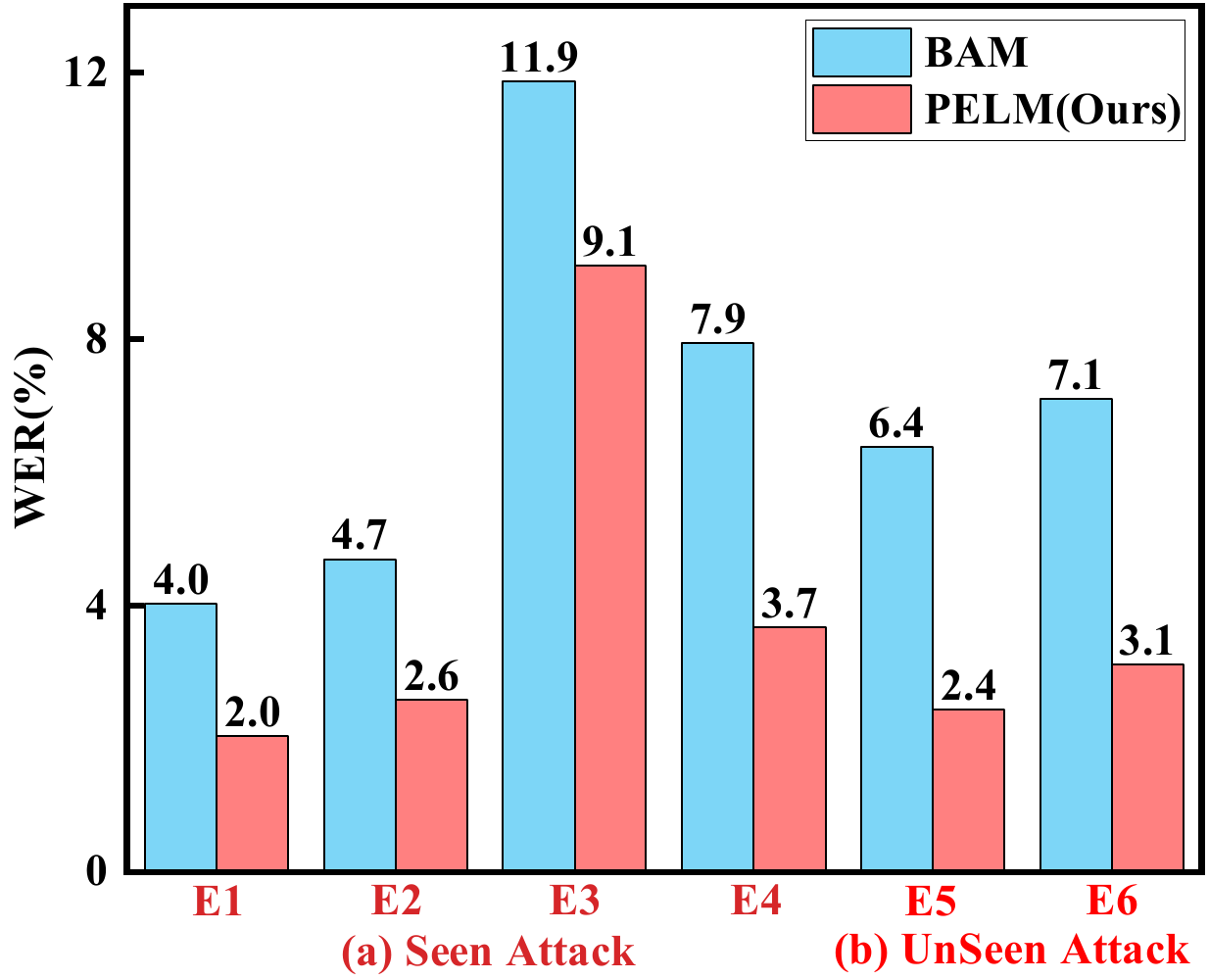}
	\caption{Comparison of content localization performance (WER) on the AiEdit dataset under seen (E1–E4) and unseen (E5–E6) editing attacks.}
	\label{fig:gen}
\end{figure}

\subsection{Generalization to unseen editing attacks}

To evaluate the generalization ability under unseen editing attacks, we conduct experiments on the AiEdit dataset under both seen (E1–E4) and unseen (E5–E6) settings. The content localization performance (WER) is shown in Fig.~6.
Under both seen and unseen settings, the proposed PELM consistently outperforms the baseline BAM, demonstrating stable performance gains. For example, on E4, PELM reduces the WER from 7.9\% to 3.7\%, corresponding to a relative improvement of approximately 53.2\%. On unseen attacks (E5 and E6), the WER is further reduced from 6.4\% and 7.1\% to 2.4\% and 3.1\%, respectively.
Overall, PELM achieves the best performance under both seen and unseen editing attacks, indicating its ability to learn generalizable editing representations.

\subsection{Ablation study}

To validate the effectiveness of each component in PELM and explore optimal configurations, we conduct comprehensive ablation studies across five dimensions: prompt design, LoRA hyperparameters, fine-tuning scope, backbone Audio LLMs, and key module components. The results are summarized in Table~\ref{tab:ablation_study}.

\noindent\textbf{Prompt strategy:} We design three prompt paradigms with increasing levels of instruction completeness to investigate the impact of instruction specificity on model performance. Prompt1 provides only basic context and input description without explicit task definition; Prompt2 further introduces task-level instructions, requiring the model to determine whether speech has been edited and identify the editing type; Prompt3 extends this by incorporating a complete task definition, explicit editing categories (Add/Delete/Modify), and strict output format constraints. The results show a clear trend that performance improves as the prompt becomes more structured and informative. On AiEdit, the WER decreases from 3.71\% (Prompt1) to 2.72\% (Prompt3), accompanied by consistent improvements in detection metrics. This indicates that Audio LLMs are highly sensitive to instruction design: clearer task definitions reduce ambiguity in reasoning, while stricter output constraints effectively guide the model to focus on acoustically relevant regions rather than generating unconstrained semantic responses.

\noindent\textbf{LoRA hyperparameters:} We analyze the impact of the low-rank dimension $r$ and scaling factor $\alpha$ in LoRA. Here, $r$ controls the dimensionality of the trainable subspace and thus determines the expressive capacity of parameter updates, while $\alpha$ scales the magnitude of these updates relative to the pre-trained weights. Experimental results show that smaller configurations (e.g., $r=16, \alpha=16$) lead to inferior performance due to limited modeling capacity, whereas increasing both $r$ and $\alpha$ consistently improves results. The best performance is achieved at $r=32, \alpha=64$, suggesting that modeling fine-grained acoustic inconsistencies requires sufficient parameter flexibility. However, further increasing the parameter scale yields diminishing returns, indicating that the model capacity has reached a saturation point for this task.

\noindent\textbf{Fine-tuning scope:} We explore three different fine-tuning strategies: (1) Full-depth LoRA (1.47\%), where LoRA is applied to all linear layers across both audio and text modules; (2) full fine-tuning of the entire audio encoder (11.19\%), allowing maximum adaptation of acoustic representations; and (3) full fine-tuning of only the last three layers of the audio encoder along with the projection layer (2.27\%), while keeping the rest under LoRA. The results show that the last strategy achieves the best trade-off between performance and efficiency. This can be attributed to the hierarchical nature of speech representations: lower layers mainly encode basic acoustic features, while higher layers capture more abstract semantic and contextual information. Since speech editing detection relies more on high-level consistency and anomaly patterns, adapting only the top layers is sufficient. In contrast, fully fine-tuning the entire encoder introduces substantial computational overhead without proportional performance gains.

\begin{table}[t]
	\centering
	\caption{Ablation Study of Three dimensions. The 'Train Param Ratio' column indicates the percentage of trainable parameters relative to the full model.}
	\label{tab:ablation_study}
	\resizebox{\columnwidth}{!}{%
		\setlength{\tabcolsep}{3pt} 
		\renewcommand{\arraystretch}{1.02}
		
		\begin{tabular}{l c cc c c cc c}
			\toprule
			\multirow{3}{*}{\textbf{Configuration}} & \multirow{3}{*}{\shortstack{\textbf{Train} \\ \textbf{Param} \\ \textbf{Ratio}}} & \multicolumn{3}{c}{\textbf{HumanEdit}} & & \multicolumn{3}{c}{\textbf{AiEdit}} \\
			\cmidrule(lr){3-5} \cmidrule(lr){7-9}
			& & \multicolumn{2}{c}{Det.} & Loc. & & \multicolumn{2}{c}{Det.} & Loc. \\
			\cmidrule(lr){3-4} \cmidrule(lr){5-5} \cmidrule(lr){7-8} \cmidrule(lr){9-9}
			& & ACC $\uparrow$ & EER $\downarrow$ & WER $\downarrow$ & & ACC $\uparrow$ & EER $\downarrow$ & WER $\downarrow$ \\
			\midrule
			
			\rowcolor{cmp}
			\multicolumn{9}{l}{\textbf{\textit{Prompt strategy: see figs~\ref{fig:prompt1}}}} \\
			
			Prompt1          & - & 99.4 & 0.56 & 9.68& & 91.48 & 12.49 & 3.71\\
			Prompt2          & - & 99.44  & 0.55 & 10.56& & 92.46 & 11.14 & 3.55\\
			Prompt3          & - & \textbf{99.62} & \textbf{0.55} & \textbf{9.17} & & \textbf{95.2} & \textbf{8.37} & \textbf{2.72} \\
			
			\midrule
			
			\rowcolor{base}
			\multicolumn{9}{l}{\textbf{\textit{LoRA hyperparams}}} \\
			~~$r$=16, $\alpha$=16 & - & 98.96 & 0.62 & 10.21& & 91.21 & 13.76 &3.66  \\
			~~$r$=16, $\alpha$=32 & - & 98.94 & 0.55 & 9.51& & 94.17 & 11.42 &3.24 \\
			~~$r$=32, $\alpha$=32 & - & 98.88 & 0.59 &9.54& & 94.13 & 12.14 &3.22 \\
			~~$r$=32, $\alpha$=64 & - & \textbf{99.62} & \textbf{0.55} & \textbf{9.17} & & \textbf{95.2} & \textbf{8.37} & \textbf{2.72} \\
			
			\midrule
			
			\rowcolor{abl}
			\multicolumn{9}{l}{\textbf{\textit{Fine-tuning scope}}} \\
			~~Full-depth LoRA   & \textbf{1.47\%}  & 99.02 & 0.56 & 9.44& & 94.35 & 10.11 &3.71 \\
			~~Audio encoder     & 11.19\% & 99.04 & 0.55 & 10.56& & 94.34 & 10.11 & 3.55\\
			~~Last 3 layers     & 2.27\%  & \textbf{99.62} & \textbf{0.55} &  \textbf{9.17}& & \textbf{95.2} & \textbf{8.37} &  \textbf{2.72}\\
			\midrule
			\rowcolor{enc}
			\multicolumn{9}{l}{\textbf{\textit{Different Audio LLMs}}} \\
			Qwen2-audio-7B  & - & 97.51  & 0.74 & 19.03& & 89.36 & 44.88 &7.24\\
			Qwen2.5-omni-7B     & -&  \textbf{99.76} & 0.64 & \textbf{8.90 }& &  \textbf{95.29} & 10.11 & 3.94\\
			Qwen2.5-omni-3B     & -&99.62 & \textbf{0.55} &  9.17& &95.20 & \textbf{8.37} &  \textbf{2.72}\\
			\midrule
			\rowcolor{rob}
			\multicolumn{9}{l}{\textbf{\textit{Ablation experiments }}} \\
			w/o Prior-enhanced  & - &  88.56 &18.77 & 19.67 &&66.87 & 35.80  &11.78 \\
			w/o Acoustic-constraint    & -   &99.30& \textbf{0.49}& 10.17& & 96.67& 9.14  &  3.75 \\
			PELM    & -& \textbf{99.62} & 0.55 &   \textbf{9.17}& & \textbf{95.20} & \textbf{8.37} &  \textbf{2.72}\\
			
			\bottomrule
		\end{tabular}%
	}
\end{table}

\noindent\textbf{Different Audio LLMs:} We further compare different backbone Audio LLMs, including Qwen2-audio-7B, Qwen2.5-omni-7B, and Qwen2.5-omni-3B. The results show that Qwen2.5-omni-3B achieves the best overall performance across both detection and localization tasks, while the larger Qwen2.5-omni-7B does not consistently outperform it and even shows slight degradation on AiEdit. Meanwhile, Qwen2-audio-7B performs significantly worse across all metrics. These observations suggest that model scale alone is not the determining factor; instead, multimodal alignment capability and pretraining design play a more critical role in speech editing detection.

\noindent\textbf{Module ablation:} Finally, we evaluate the contributions of the prior-enhanced design and the acoustic consistency constraint. Removing the prior leads to a significant performance drop, with WER on AiEdit increasing from 2.72\% to 11.78\%, indicating that semantic reasoning alone is insufficient for identifying fine-grained editing artifacts. The word-level probabilistic prior provides explicit acoustic cues that help constrain the reasoning boundary of the model. Similarly, removing the acoustic consistency constraint degrades performance (WER increases to 3.75\%), demonstrating its role in enhancing sensitivity to local anomalies by modeling feature-space distributions. Overall, the prior and the consistency constraint operate in a complementary manner: the prior guides the model toward potential anomalous regions, while the consistency constraint enforces discriminative feature structures, leading to the best performance when combined.

\begin{table}[t]
	\centering
	\caption{Cross-dataset evaluation results on HumanEdit and AiEdit. H2A denotes training on HumanEdit and testing on AiEdit, while A2H denotes training on AiEdit and testing on HumanEdit.}
	\label{tab:cross_domain}
		\setlength{\tabcolsep}{2pt} 
		\renewcommand{\arraystretch}{1.02}
		
		\begin{tabular}{l cc c cc}
			\toprule
			\multirow{3}{*}{\textbf{Model}} & \multicolumn{2}{c}{\textbf{H2A}} & & \multicolumn{2}{c}{\textbf{A2H}} \\
			\cmidrule(lr){2-3} \cmidrule(lr){5-6}
			& Det. & Loc. & & Det. & Loc. \\
			\cmidrule(lr){2-2} \cmidrule(lr){3-3} \cmidrule(lr){5-5} \cmidrule(lr){6-6}
			& EER $\downarrow$ & WER $\downarrow$ & & EER $\downarrow$ & WER $\downarrow$ \\
			\midrule
			
			TDL~\cite{tdl}   & 41.36 & 55.30 & & 40.93 & 48.61 \\
			CFPRF~\cite{cfprf} & 51.3 & 41.25 & & 60.49 & 50.31 \\
			AGO~\cite{ago}   & 40.14 & 47.58 & & 36.04 & 49.86 \\
			BAM~\cite{bam}   & 38.13 & 56.92 & & 36.72 & 62.26 \\
			\textbf{PELM (Ours)} & \textbf{19.08} &\textbf{13.35}  &&\textbf{1.02} & \textbf{41.32} \\
			
			\bottomrule
		\end{tabular}%
\end{table}

\subsection{Cross-domain Generalization}

To evaluate cross-domain generalization, we conduct cross-dataset experiments between HumanEdit and AiEdit. As shown in Table~6, existing methods suffer from significant performance degradation under domain shifts, while PELM consistently outperforms all baselines in both detection (EER) and localization (WER).
Specifically, under the H2A setting, PELM achieves substantial improvements, demonstrating strong adaptability to more realistic neural editing scenarios. Under the A2H setting, PELM achieves an extremely low EER of 1.02\%, indicating robust detection capability. However, the localization performance is relatively weaker.
This phenomenon may be attributed to the nature of HumanEdit, which is primarily based on frame-level splicing or replacement. Such operations tend to disrupt semantic continuity, making localization more dependent on low-level acoustic artifacts rather than semantic reasoning.

\section{Conclusion}

In this paper, we address the problem of speech editing detection and content localization under increasingly realistic and diverse editing scenarios. We first construct a bilingual speech editing dataset, AiEdit, which covers multiple editing operations including addition, deletion, and modification, providing a more comprehensive benchmark for evaluating modern editing threats.
To overcome the limitations of conventional frame-level methods, we propose a prior-enhanced audio large language model (PELM) that formulates detection and localization as a unified text generation task. By incorporating word-level acoustic priors and an acoustic consistency-aware loss, the proposed framework effectively balances semantic reasoning and low-level acoustic modeling, enabling robust detection of subtle editing artifacts.
Extensive experiments demonstrate that PELM achieves superior performance across both detection and localization tasks, and exhibits strong cross-domain generalization ability under different editing paradigms. These results indicate that the proposed method can capture more domain-invariant spoofing cues and provide a promising direction for future research on realistic speech editing detection.

\bibliography{example_paper}
\bibliographystyle{icml2026}
\end{document}